\documentclass[aps,prl,twocolumn,superscriptaddress,showpacs]{revtex4}

\usepackage{graphicx}
\usepackage{dcolumn}
\usepackage{amsmath}

\begin{document}

\title{Ferroelectric switched all-metallic-oxide $p$-$n$ junctions}
\author{J. Yuan, H. Wu, L. Zhao, K. Jin, L. X. Cao, B. Y. Zhu, S. J. Zhu, J. P. Zhong, J. Miao, H. Yang, B. Xu, X. Y. Qi, Y. Han, X. G. Qiu, X. F. Duan and B. R. Zhao}\email{brzhao@aphy.iphy.ac.cn}
\affiliation{National Laboratory for Superconductivity, Institute
of Physics, and Beijing National Laboratory for Condensed Matter
Physics, Chinese Academy of Sciences, Beijing 100080, China}

\date{\today}

\begin{abstract}
We report the first formation of the metallic $p$-$n$ junctions,
the ferroelectric (Ba,Sr)TiO$_3$ (BST) switched optimally
electron-doped ($n$-type) metallic T'-phase superconductor,
(La,Ce)$_2$CuO$_4$ (LCCO), and hole-doped ($p$-type) metallic CMR
manganite (La,Sr)MnO$_3$ (LSMO) junctions. In contrast with the
previous semiconductor $p$-$n$ ($p$-$I$-$n$) junctions which are
switched by the built-in field $V_0$, the present metallic oxides
$p$-$I$-$n$ junctions are switched by double barrier fields, the
built-in field $V_0$, and the ferroelectric reversed polarized
field $V_{rp}$, both take together to lead the junctions to
possess definite parameters, such as definite negligible reversed
current ($10^{-9}$ A), large breakdown voltage ($>$7 V), and
ultrahigh rectification ($>2\times10^4$) in the bias voltage 1.2 V
to 2.0 V and temperature range from 5 to over 300 K. The related
transport feature, barrier size effect, and temperature effect are
also observed and defined.
\end{abstract}

\pacs{85.50.-n, 73.40.-c, 73.40.Ei, 74.78.Fk}

\maketitle

Following the semiconductor technology, for which the $p$-$n$
junctions are the basic devices, the oxide electronics, such as
the high-$T_C$ superconductor (HTSC) electronics, spintronics
based on the colossal magnetoresistance (CMR) manganites, and so
on, the oxide $p$-$n$ ($p$-$I$-$n$) junctions (oxide diodes) also
are the basic devices, and have been studied widely. One route, a
hole-doped oxide(s), currently, the $p$-type CMR manganite(s) is
used, which is deposited on the $n$-type semiconductor oxide to
form the $p$-$n$ junctions \cite{Tanaka_88_027204_PRL,
Tiwari_APL_83_1773, Hu_APL_83_1867,
Lu_APL_86_2502,Xi_APL_61_2353,Ramadan_JAP_99_43906}. The other
route, the oxide insulator is used to sandwich the semiconductor
$p$-type oxide and semiconductor $n$-type oxide (or $n$-type pure
semiconductor) to prepare the $p$-$I$-$n$ junctions
\cite{Sugiura_JAP_90_187,Mitra_APL_79_2408,Lang_APL_87_3502}. It
must be noted that the semiconductor functional oxides are usually
due to the under doping ,the optimally doped $p$- and $n$-type
oxide functional materials are metallic phase oxides with their
maximum functions. Therefore, for really developing the oxide
electronics and making matchable oxide electronic circuits,
exploiting a way to prepare entire new oxide $p$-$n$ junctions
based on the optimally doped metallic functional oxides is an
imperative topic. It is known that the origin of the rectifying
function of the conventional semiconductor $p$-$n$ junctions is
the formation of the potential barrier, called the built-in field
$V_0$ in the junction interface based on the energy band structure
of $p$- and $n$- type semiconductors\cite{Streetman}. For the
metallic oxide $p$-$n$ junctions, the energy band structure in the
interface is entirely different. Because in the metallic state,
the Fermi level lies within the conduction band for both sides of
the junction, the built-in field $V_0$ can not be formed by
$p$-region and $n$-region themselves, since no energy band bending
can form, no rectifying function can be induced, so no metallic
$p$-$n$ junctions are developed so far. Therefore setting up the
barrier potential in the interface of the metallic $p$-$n$
junction, i.e., making it into the metallic $p$-$I$-$n$ junction
is a crucial topic. For this the key is the selection of the
barrier materials. It has been shown that the conventional
insulator, such as SrTiO$_3$ is not the energetic material to
build the high and wide enough barrier to get the stable
rectifying function even for the semiconductor oxide $p$-$I$-$n$
junctions\cite{Sugiura_JAP_90_187,Mitra_APL_79_2408,Lang_APL_87_3502}.
In order to solve this problem, the very possibility is to use the
perovskite structural ferroelectric as the barrier to form the
metallic $p$-$I$-$n$ junction, since in the case of zero applied
voltage, the ferroelectric can act as an usual insulator, in which
a built-in field $V_0$ can from as in the conventional $p$-$I$-$n$
junctions; under the external bias voltage, the ferroelectric
possesses reversible polarization feature due to its spontaneous
polarization nature, resulting in the formation of a reversed
polarized field $V_{rp}$ in the junction. Then in contrast to the
conventional insulator barrier, the ferroelectric barrier produces
double barrier fields, $V_0$ and $V_{rp}$, which possibly induce
the rectifying function for the metallic oxide $p$-$I$-$n$
junctions. (Very recently, it is supposed that the ferroelectric
reversed polarized field can drive the tunneling junction to have
special function\cite{Tsymbal}).

In the present work, we developed a kind of metallic $p$-$I$-$n$
junctions by using the ferroelectric (Ba$_{0.5}$Sr$_{0.5}$)TiO$_3$
(BST) as the barrier layer for its high dielectric constant
($>$1000) and high Curie temperature ($\geq$300 K), the metallic
phase ferromagnetic Sr-doped LaMnO$_3$,
(La$_{0.67}$Sr$_{0.33}$)MnO$_3$ (LSMO) with Curie temperature
$\sim$350K as the $p$-region, and the optimally Ce-doped
La$_2$CuO$_4$, T'-phase (La$_{1.89}$Ce$_{0.11}$)CuO$_4$ (LCCO),
which possesses the highest superconducting transition temperature
(30 K) in the T'-phase electron-doped cuprate superconductor
family \cite{Dagotto,Naito,Zhao} as the $n$-region. Both LSMO and
LCCO possess the carrier concentrations of $10^{21}$ cm$^{-3}$
order \cite{Naito,Wu,Zhao,Salamon}. The junctions possess the
definite negligible reversed current, large reversed voltage and
ultrahigh rectification in the low bias voltage and the
temperature range of 5 to $>$300 K .

The $p$-$I$-$n$ junctions of LCCO/BST/LSMO are prepared by pulsed
Laser deposition method on (001) SrTiO$_3$ (STO) substrate
basically following the growth conditions of LCCO, BST, and
LSMO\cite{Wu,Chen,Zhang}. The thickness of each layer is
controlled by the number of the pulse of the laser beam. The
junctions (with both LSMO and LCCO in $\sim$100 nm) with BST with
thickness 5, 15, 25, 50, and 100 nm are designed. The whole
junction is in perfect epitaxial growth along (001) direction, no
structure defects are observed (Fig.2(a)). The intrinsic evidence
of ferroelectricity, the $p$-$V$ hysteresis loops are observed
obviously for BST$\geq$20 nm in the junction (the inset of Fig. 3
(a)).

\begin{figure}
\resizebox{0.45\textwidth}{!}{\includegraphics{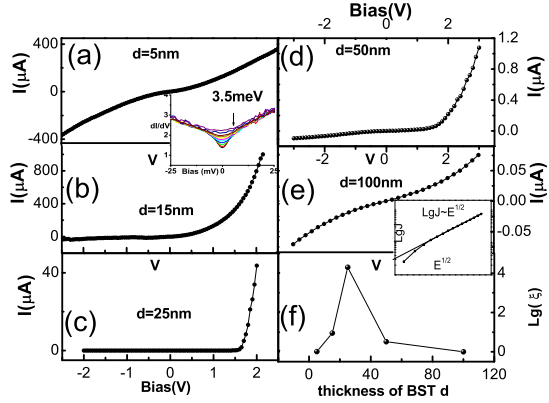}}
\caption{The $I$-$V$ curves of the junctions with various thickness
of BST at 5 K. (a) The thickness of BST $\sim$5 nm which is much
smaller than the critical thickness for
ferroelectricity\cite{Chen,Junquera,Gerra,Fong}, the junction works
as the conventional single particle superconducting tunneling
junction, it gives an energy gap of 3.5 meV for the LCCO (inset of
(a)). (b-d) With increasing the thickness, the asymmetry enhances.
$p$-$I$-$n$ junction forms, the largest rectification ($>10^4$) is
obtained for BST$\sim$25 nm (c). (e) The BST is thick as large as
100nm, it acts as a block ferroelectric layer to limit the leakage
current between p- and n-regions, the current shows Schottky model
in the low electric field(see the inset) and space-charge-limited
model in the high electric field (not shown in the figure); (f)The
role of the thickness (ferroelectric) of BST layer on rectifying
function in the present junction, the critical thickness for
ferroelectricity of BST is $\sim$10 nm, the size for the full
ferroelectricity of BST is 20-25 nm.\label{ivd}}
\end{figure}

In such nanometer scale of BST, the thickness directly reflects
the ferroelectricity, so the shape of $I$-$V$ curves of the
junction directly depends on the thickness of BST. Fig.1 (a-e)
show the $I$-$V$ curves of the LCCO/BST/LSMO junctions with
various thickness of BST. It is clearly indicated that the
ferroelectric plays the key role on the rectifying function of
$p$-$I$-$n$ junction intrinsically based on $V_0$ and $V_{rp}$,
the $V_0$ always directs from LCCO (n-region) to LSMO (p-region)
in the junction\cite{Streetman}, and blocks the motion of
electrons from the LCCO to LSMO. When we take the external bias
voltage to be positive in LCCO relative to LSMO, the reversed
polarized field $V_{rp}$ of ferroelectric BST must direct from
LSMO to LCCO, and suppresses the $V_0$, resulting in the
occurrence of the forward current from LCCO to LSMO. In the
reversed bias case, the direction of $V_{rp}$ of BST is the same
as that of $V_0$, then both $V_{rp}$ and $V_0$ take together to
largely block the reversed current. So we can conclude that the
$V_{rp}-V_0$ promotes the forward current, and the $V_{rp}+V_0$
blocks the reversed current of the junction. As shown in Fig.2. By
which, we present the high resolution TEM image of the section ,
the schematic of the energy band structure in the interface, and
the schematic of mechanism (and design) of such metallic
$p$-$I$-$n$ junction. The role of the ferroelectric BST is clearly
presented.

\begin {figure}
\resizebox{0.45\textwidth}{!}{\includegraphics{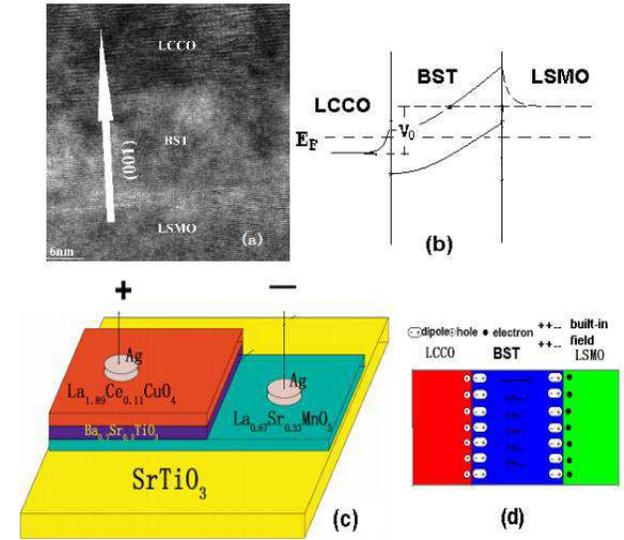}}
\caption{(a) The high resolution TEM (HRTEM) image of the cross
section of the junction, shows the perfect epitaxial growth of
(001)LSMO/(001)BST/(001)LCCO on (001) SrTiO$_3$ substrate, no defect
are observed. (b) The schematic of the energy band of the
$p$-$I$-$n$ junction, the energy band bending occurs in the BST due
to the formation of $V_0$ within it, no energy band bending occurs
within the metallic $p$- and $n$-regions, but there is a sharp
bending in the interface due to the formation of reversed polarized
field $V_{rp}$ of BST under the bias voltage (there also should be a
Schottky barrier in the interface, but it is smaller than $V_{rp}$).
(c) The sketch of the junction. In order to get the forward current
(electrons from $n$-region to $p$-region) in the forward bias, it is
designed to take the external bias to be positive on $n$-region
(LCCO) relative to the $p$-region (LSMO). From (d) within the BST,
the whole barrier potential is the series of $V_{rp}$ and $V_0$ in
several zones, sum up, the $V_{rp}-V_0$ promotes the forward
current, and the $V_{rp}+V_0$ blocks the reversed current (as in the
present case of Fig.2(d)).\label{sketch}}
\end{figure}

To further understand the current feature in such $p$-$I$-$n$
junctions, we make examination of the  relation of the forward
current and bias voltage for the junction with BST$\sim$25nm as
shown in Fig.1(c). It is shown in Fig.3 that below a bias voltage
which is determined by using the $\ln I \sim V^{\frac{1}{2}}$, we
obtained the relation of $\ln I\propto V^{\frac{1}{2}}$, i.e., the
current obeys the Schottky model\cite{Yang}, this critical bias
voltage can be considered to be the Schottky barrier $V_s$, which
is descended with increasing temperature (the inset of Fig.3),
from 1.15 V at 5 K to 0.98 V at 100 K, i.e., below $V_s$, the
current is limited by the Schottky emission though $V_{rp}$ and
$V_0$ exist. In the higher bias voltage, the current deviates from
the Schottky model. When the bias voltage is increased till $>$ 2
V, the current seems still not to approach the
Space-charge-limited model caused in heterostructure with
conventional size of BST\cite{Yang}, the current is completely
dominated by $V_{rp}$ and $V_0$. So that , the BST with thickness
of 20$\sim$25 nm in such junction may not currently form
space-charge-limited type current unless in the higher bias
voltage region. The $p$-$I$-$n$ junction with BST$\sim$20-25 nm is
quite different from that with BST with thickness $>$50 nm, for
which, as shown above the current respectively obeys the Schottky
model in the low bias voltage, and Space-charge-limited model in
the higher bias voltage. Therefore, based on this aspect, we may
make more complete conclusion that for configuring the present
$p$-$I$-$n$ junction, the BST layer can not be too thin since
$V_{rp}$ and $V_0$ can not form when the BST is thinner than a
limitation of thickness ($<$10 nm); the BST can not be too thick
since the Schottky emission and Space-charge-limited model will
dominate the current of the junction when BST is thicker than a
limitation of thickness ($>$50 nm). To form the optimal
LCCO/BST/LSMO $p$-$I$-$n$ junctions, from the experimental data
indicated above, the thickness of BST should be 20-25 nm, for
which $V_0$ and $V_{rp}$ can form and can play dominated roles in
the suitable bias voltage region.

\begin {figure}[!]  
\begin{center}
\includegraphics*[bb=79 420 472 719, width=7cm, clip]{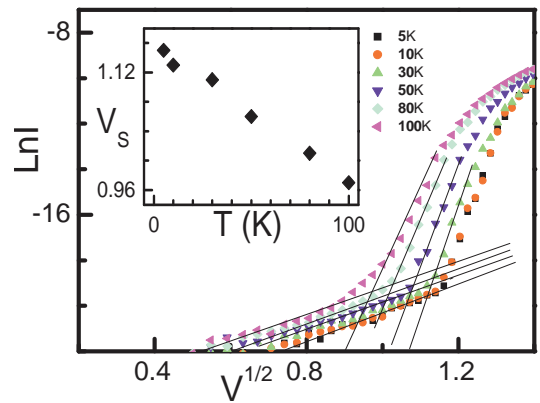}
\caption{The relation of $\ln I$ and $V^{1/2}$ for LCCO/BST/LSMO
with BST$\sim$25 nm (from Fig.1(c)) at various temperature. In the
low electric field region the current obeys the Schottky model; with
increasing bias voltage, the current is just dominated by $V_{rp}$
and $V_0$.\label{lnJ_$V^{1/2}$}}\end{center}
\end{figure}

In order to reveal the ferroelectric role on the LCCO/BST/LSMO
junction further, we investigated the temperature dependence of
the rectifying function. Fig. 4(a) shows the temperature
dependence of the $I$-$V$ curve shape of the LCCO/BST/LSMO
junction with BST $\sim$ 25 nm in the temperature range from 5 K
to 300 K. Contrasting with the SrTiO$_3$ switched semiconductor
oxide $p$-$I$-$n$ junctions \cite{Sugiura_JAP_90_187}, it clearly
indicates that the shape of these $I$-$V$ curves of the present
$p$-$I$-$n$ junctions is basically independent of the temperature.
But the temperature of rectifying function is needed to be
investigated carefully for the practical applications. For this
kind of $p$-$I$-$n$ junctions, the bias voltage is of 1-2 volts,
which is much larger than the superconducting energy gap, so these
$p$-$I$-$n$ junctions all work on the normal state transport of
LCCO, which  doesn't show strong temperature dependence.
Therefore, the temperature dependence of the $I$-$V$ curves of the
junction should be basically caused by the temperature dependence
of the ferroelectricity of BST. Intrinsically,as indicated above,
$V_{rp}-V_0$ switches the forward current, $V_{rp}+V_0$ blocks the
reversed current. So the temperature dependence of the $I$-$V$
curve shape should originate from the temperature dependence of
$V_0$ and $V_{rp}$. To define this, we characterize $V_0$ and
$V_{rp}$ in the following two critical points: the forward bias at
which the forward current starts to occur (the criterion is
$1\times10^{-6}$ A at 5 K) is the value of $V_{rp}-V_0$ ; the
reversed bias at which the reversed current starts to occur(the
criterion is $1\times 10^{-7}$ A at 5 K) is the value of sum of
$V_{rp}+V_0$. Such two points can critically reveal the
temperature dependence of $V_0$ and $V_{rp}$. At 5K, the forward
current starts to occur at the forward bias voltage of 1.5V; the
reversed current starts to occur at -7 V, and can be estimated by
using the relation that $V_{rp}-V_0$= 1.5 V, $V_{rp}+V_0$= 7 V
then $V_{rp}$= 4.25 V, $V_0$= 2.75 V. By this way, every couple of
$V_0$ and $V_{rp}$ at each temperature can be defined and
concluded in Fig. 4(b). On the other hand, the rectification of
the junction with BST$\sim$25 nm at various temperatures also can
be defined. At 5K, the forward current at the forward bias of 2V
is $4.5\times10^{-5}$ A, the reversed current at the reversed bias
(-2 V) is $2.0\times10^{-9}$ A, the rectification, defined as
$\xi$=forward current at +2 V/reversed current at -2 V is
$2.2\times10^4$. When the temperature is increased from 10 K to 80
K, the forward current at +2 V increased from 4.6$\times10^{-5}$ A
to 7.6$\times10^{-5}$ A, the reversed current increase from
2$\times 10^{-9}$ A to 2.5$\times10^{-9}$ A. the rectification
changes from 2.3$\times10^4$ to 3$\times10^4$. At 100 K, the
rectification has a rather larger increase, up to 3.8$\times10^4$.
At 300 K, the forward current is more than one order larger than
that at 100 K (the reversed current is in $10^{-8}$ A order) the
rectification keeps in a high value of $10^4$. So the
rectification of the present $p$-$I$-$n$ junctions shows obvious
positive temperature dependence in the temperature range up to
$>$300 K (the inset of Fig.4(b)) and the main contributor for this
is $V_{rp}$. The big change of $V_{rp}$ and $V_0$ of the junction
at $\sim$100 K may come from a phase transition of BST in the
junction with such nanometer size. This should be examined
further. But this big change does not make a hindrance on the
rectifying function of such $p$-$I$-$n$ junction. Therefore, the
ultrahigh rectification ($>10^4$) and its positive temperature
dependence makes the present metallic $p$-$I$-$n$ junction to have
rich power to satisfy the requirements of the practical
applications. It also indicates that the BST with wide forbidden
band of $\sim$3.5 eV \cite{Dawber} is a very suitable material to
make oxide metallic $p$-$I$-$n$ junctions.

\begin {figure}
\resizebox{0.45\textwidth}{!}{\includegraphics{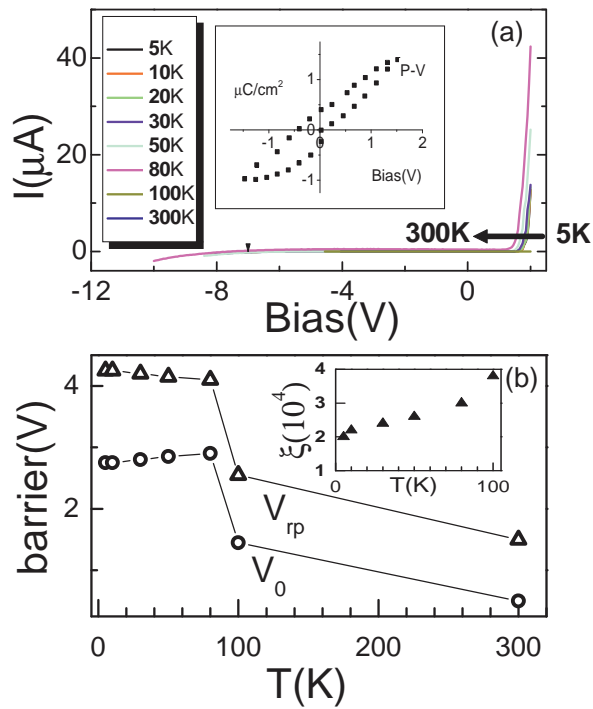}}
\caption{(a) The $I$-$V$ curves at various temperatures for the
LCCO/BST/LSMO $p$-$I$-$n$ junction with BST$\sim$25 nm shown in
Fig.1(c). it basically shows the temperature independence of the
shape of the $I$-$V$ curves , the inset is the $p$-$V$ hysteresis
loop of BST with thickness $\sim$ 25 nm examined at 300 K. (b) The
$V_0$ and $V_{rp}$ are estimated from (a), which have almost the
same temperature dependence, but the $V_{rp}/V_0$ increases with
increasing temperature, this indicates that the positive temperature
dependence of rectification of the $p$-$I$-$n$ junction (shown in
the inset) is mainly contributed by ferroelectric reversed polarized
field.\label{i_v_t}}
\end{figure}

In conclusion the metallic oxide LCCO/BST/LSMO $p$-$I$-$n$
junctions consisting of three main kinds of  functional oxides are
successfully made for the first time. Due to the switch role of
ferroelectric critically realized by double barrier fields $V_0$
and $V_{rp}$, this kind of $p$-$I$-$n$ junctions with
BST$\sim$20-25 nm show definite unique functions. Contrasting to
the standard semiconductor $p$-$n$ junctions, for which there is
no forward bias for starting the forward current, and there is a
finite reversed current\cite{Streetman}, the present metallic
oxide $p$-$I$-$n$ junctions possess definite forward bias voltage
for starting the forward current and possess the definite
negligible reversed current, so they are the real oxide diodes
with strict switch function; contrasting to the semiconductor
oxide $p$-$n$ ($p$-$I$-$n$) junctions for which the functions are
to be defined, the present metallic oxide $p$-$I$-$n$ junctions
possess definite ultrahigh rectifying function with positive
temperature coefficient, for which the $I$-$V$ curve shape is
temperature independent. All these functions of the metallic oxide
$p$-$I$-$n$ junctions maintain in the temperature range of 5 to
$>$300 K, making them to be real basic devices in the field of
oxide electronics.
\begin{acknowledgments}
We thank  Prof. M. Tachiki,  Prof. X. Hu, $\&$  Prof. X. X. Xi, for
their helpful discussions. This work is supported by grants from the
State Key Program for Basic Research of China (No. 2004CB619004-1)
and the National Natural Science Foundation (No. 10474121).
\end{acknowledgments}

\end{document}